\documentclass[fleqn,usenatbib]{mnras}
\usepackage{graphicx}
\usepackage{txfonts}
\usepackage[T1]{fontenc}
\usepackage{threeparttable}
\usepackage{xcolor}
\usepackage{comment}
\usepackage{url}

\def\mum    {$\mu$m}
\def\kms    {km\,s$^{-1}$}

\graphicspath{{New/}}
\let\biblio\bibliography
\let\bibsty\bibliographystyle
\renewcommand{\bibliography}[1]{\expandafter\biblio{New/#1}}
\renewcommand{\bibliographystyle}[1]{\expandafter\bibsty{New/#1}}

\DeclareRobustCommand{\VAN}[3]{#2}
\let\VANthebibliography\thebibliography
\def\thebibliography{\DeclareRobustCommand{\VAN}[3]{##3}\VANthebibliography}

\title[Discovery of an optically-dark galaxy at $z$ = 3.4]{Serendipitous Discovery of an Optically-Dark Ultra-Luminous Infrared Galaxy at $z$ = 3.4}

\author[N. H. Hayatsu et al.]{Natsuki\,H.~Hayatsu,$^{1}$\thanks{E-mail: natsuki.hayatsu@aoni.waseda.jp}
Zhi-Yu~Zhang,$^{2,3}$
R.\,J.~Ivison,$^{4}$
Chao-Wei~Tsai,$^{5,6,7}$
Ping~Zhou,$^{2,3}$
\newauthor
Katsuya~Okoshi,$^{8}$
Chentao~Yang,$^{9}$
Yuri~Nishimura,$^{10,11,12,13}$
Kotaro~Kohno,$^{10,14}$
\newauthor
Nobunari~Kashikawa,$^{15}$
Masahiro~Nagashima,$^{16}$
Junfeng~Wang,$^{17}$
and Denis~Burgarella$^{18}$
\\
$^{1}$Waseda Research Institute for Science and Engineering, 3-4-1 Okubo, Shinjuku, Tokyo 169-8555, Japan\\
$^{2}$School of Astronomy and Space Science, Nanjing University, Nanjing 210023, People’s Republic of China\\
$^{3}$Key Laboratory of Modern Astronomy and Astrophysics (Nanjing University), Ministry of Education, Nanjing 210023, People’s Republic of China\\
$^{4}$European Southern Observatory, Karl-Schwarzschild-Strasse~2, D-85748 Garching, Germany\\
$^{5}$National Astronomical Observatories, Chinese Academy of Sciences, 20A Datun Road, Beijing 100101, China\\
$^{6}$Institute for Frontiers in Astronomy and Astrophysics, Beijing Normal University,  Beijing 102206, China\\
$^{7}$School of Astronomy and Space Science, University of Chinese Academy of Sciences, Beijing 100049, China\\
$^{8}$Institute of Arts and Sciences, Tokyo University of Science, 6-3-1, Niijyuku, Katsushika, Tokyo 125-8585, Japan\\
$^{9}$
Department of Space, Earth and Environment, Onsala Space Observatory, Chalmers University of Technology, 439 92 Onsala, Sweden\\
$^{10}$Institute of Astronomy, School of Science, The University of Tokyo, 2-21-1, Osawa, Mitaka, Tokyo 181-0015, Japan\\
$^{11}$ Institute of Pure and Applied Sciences, University of Tsukuba, 1-1-1 Tennodai, Tsukuba, Ibaraki, 305-8577, Japan\\
$^{12}$ Tomonaga Center for the History of the Universe, University of Tsukuba, 1-1-1 Tennodai, Tsukuba, Ibaraki, 305-8577, Japan\\
$^{13}$ Tsukuba Institute for Advanced Research (TIAR), University of Tsukuba, 1-1-1 Tennodai, Tsukuba, Ibaraki, 305-8577, Japan\\
$^{14}$Research Center for the Early Universe, School of Science, The University of Tokyo, 7-3-1 Hongo, Bunkyo, Tokyo 113-0033, Japan\\
$^{15}$Department of Astronomy, School of Science, The University of Tokyo, 7-3-1 Hongo, Bunkyo-ku, Tokyo 113-0033, Japan\\
$^{16}$Faculty of Education, Bunkyo University, 3337 Minami-Ogishima, Koshigaya-shi, Saitama 343-8511, Japan\\
$^{17}$Department of Astronomy, Xiamen University, Xiamen, Fujian 361005, People’s Republic of China\\
$^{18}$Aix Marseille Univ, CNRS, CNES, LAM, Marseille, France
}

\date{Accepted XXX. Received YYY; in original form ZZZ}
\pubyear{2025}

\begin{document}
\label{firstpage}
\pagerange{\pageref{firstpage}--\pageref{lastpage}}
\maketitle

\begin{abstract}
Dusty, submillimeter-selected galaxies without optical counterparts contribute a non-negligible fraction of the star formation in the early universe. However, such a population is difficult to detect through classical optical/UV-based surveys.
We report the serendipitous discovery of such an optically dark galaxy, behind the quadruply-lensed $z=2.56$ quasar, H1413+117, offset to the north by 6\arcsec. From $^{12}$CO $J=4$--3, $J=6$--5, and part of the $J=13$--12 transitions, which all spatially coincide with a compact
submillimeter continuum emission, we determine an unambiguous spectroscopic redshift,
$z=3.386\pm 0.005$. This galaxy has a molecular mass $M_{\rm mol} \sim 10^{11}$ M$_\odot$ and a black hole mass $M_{\rm BH} \sim 10^{8}$ M$_\odot$, estimated from $^{12}$CO $J=4$--3 and archival {\it Chandra} X-ray data ($L_{\rm 2-10,keV} \sim 4 \times 10^{44}$\,erg\,s$^{-1}$),
respectively.  We also estimate a total infrared luminosity of $L_{\rm FIR} = (2.8\pm{2.3}) \times 10^{12}$ L$_\odot$ and a stellar mass of $M_* \lesssim 10^{11}$ M$_{\odot}$, from spectral energy distribution fitting.  
According to these simple mass estimations, this gas-rich and X-ray bright galaxy might be in a transition phase from starburst to quasar offering a unique case for studying galaxy-black hole co-evolution under extremely dusty conditions.
\end{abstract}



\section{Introduction} \label{sec:intro}
The most active galaxies at high redshift often harbour large amounts of molecular gas, stellar populations, and supermassive black holes (SMBHs) and they are expected to evolve into massive elliptical galaxies in the local universe \citep[e.g.,][]{thomas2010, toft2014}. 
Such systems are usually starbursts that are luminous at far-infrared (FIR) wavelengths and faint at ultraviolet (UV) wavelengths because of dust obscuration. 
The obscuring material absorbs UV and optical photons emitted from young stars and the accretion disks of supermassive black holes; i.e., the active galactic nucleus (AGN), and re-emits at IR wavelengths. This produces a population of UV/optically dark galaxies that shine at IR and radio wavelengths \citep[e.g.,][]{wang2019}.
The most luminous quasars (QSOs)/AGNs at the peak epoch for galaxy and BH growth, $z\approx 2$, are also the most dust-reddened, often or always coincident with merging events \citep[e.g.,][]{glikman2015}.

The mid-IR all-sky survey undertaken by {\it WISE} found a population of dust-obscured AGNs, which dominate the global spectral energy distributions
(SEDs) of their host galaxies and can reach bolometric luminosities of $\approx 10^{13}$--$10^{14}$\,L$_\odot$ \citep[Hyper- or Extremely Luminous IR Galaxy;][]{eisenhardt2012,wu2012,tsai2015}.  
Their extraordinary mid-IR luminosity is due to hot dust, with little rest-frame optical emission due to the high extinction \citep[2 $\lesssim E(B-V ) \lesssim$ 25;][]{assef2015}.
Near-IR (NIR) spectroscopic studies suggest that SMBH masses can exceed 10$^9$\,M$_\odot$, with an accretion rate close to the Eddington limit \citep{wu2018,tsai2018}.  
This population of hot, dust-obscured galaxies \citep[HotDOGs;][]{wu2012} resides at the massive end of the black hole -- stellar mass relation \citep{magorrian1998,assef2015,wu2018}. 

The predecessors of the most massive galaxies are also expected to be dust-obscured starburst/AGN host galaxies, which are likely triggered by major mergers that boost star formation and AGN activity, which then blow out the surrounding dust and gas via AGN outflows and stellar winds~\citep{sanders1988, hopkins2008, tamura2010, diaz2018}.  
Although large FIR and submillimeter (submm) surveys, e.g., H-ATLAS and the SPT survey, have discovered submm galaxies (SMGs) without optical counterparts \citep{eales2010,greve2012} --- indeed, \citet{smail2021} point out that this population is rediscovered regularly --- it is extremely resource intensive and time consuming to determine their redshifts and intrinsic properties in blind line searches \citep[e.g.][]{weiss2009,walter2012}, unless the survey is designed highly efficiently using wideband receivers and/or targeting clustering fields \citep{weiss2013, neri2020}.  

In this paper, we report the serendipitous discovery of an optical and NIR dark galaxy, which has a small angular separation from a lensed QSO known as the Cloverleaf at $z = 2.56$ \citep{magain1988}.  In \S\ref{sec:ALMAobs}, we describe ALMA observations and data reduction, and in \S\ref{sec:analysis} we present the serendipitous detection of line emission in the ALMA data and multi-wavelength data including new imaging from {\it Spitzer}.  In \S\ref{sec:result}, we discuss the determination of the spectroscopic redshift and molecular and BH mass estimations.  In \S\ref{sec:summary}, we summarise our result and discuss future prospects. 

Throughout this paper, we adopt a standard $\Lambda$-CDM cosmological model with the following parameters: $\Omega_{\rm m}$\,=\,0.30, $\Omega_{\Lambda}$\,=\,0.70 and the Hubble constant, $H_{0}$\,=\,70\,\kms Mpc$^{-1}$. 
\begin{figure}
	\centering
	\includegraphics[trim= 0 0 0 0, width=70mm]{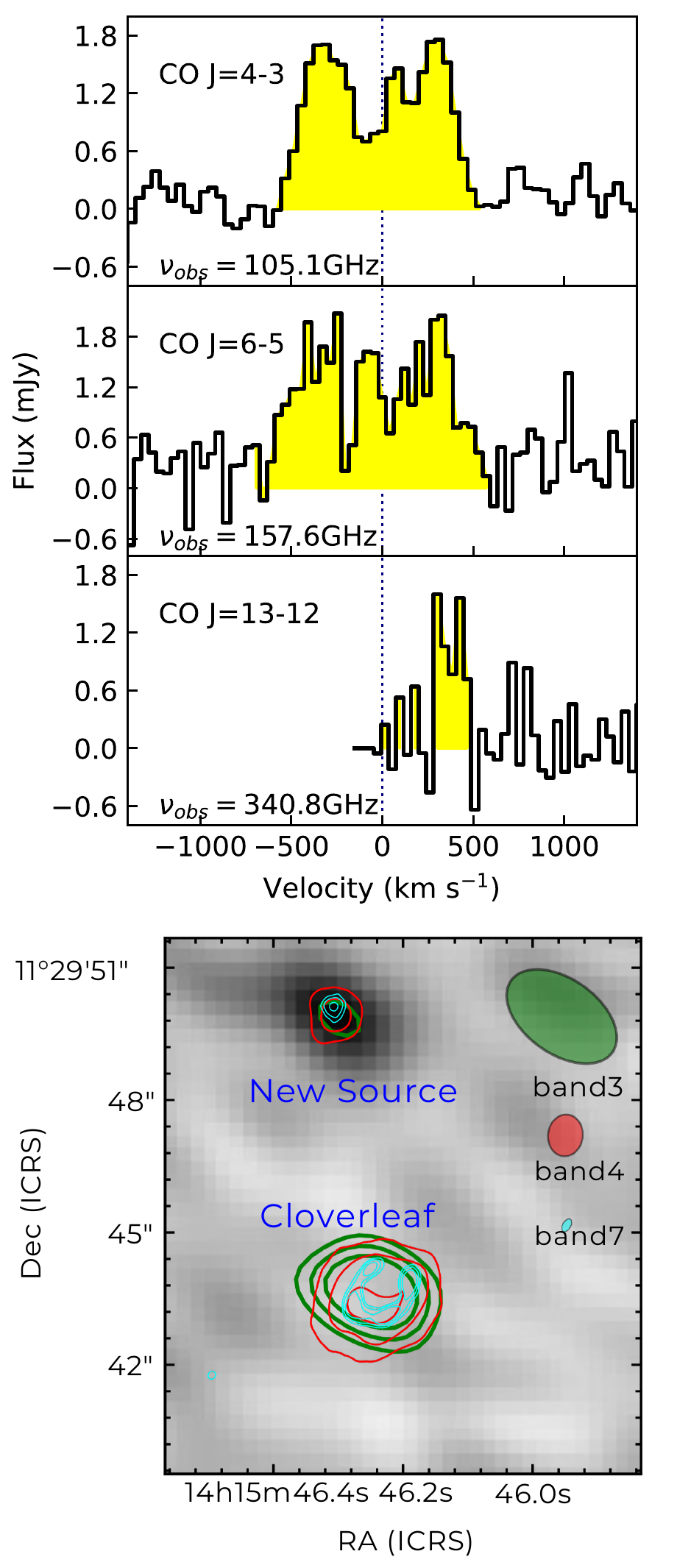}
        \caption{
         CO Spectra, submm continuum, and CO moment-0 maps of the newly discovered source at $6''$ north-east to the Cloverleaf QSO.
        {\it Top:} CO spectra $J$ = 4--3, 6--5, and 13--12 from the detected source 
        at $\sim$40--45 \kms resolution.  We label the line emission velocity range in yellow.  
        The vertical line represents $v_{\rm sys}$ = 0. The CO $J$=13-12 spectrum covers only part of the velocity range at the edge of the spectral window. 
        {\it Bottom:} Moment-0 map of the CO $J$=4--3 line detected in ALMA Band\,3, shown with grayscales. The green contours represent the Band\,3 continuum data at levels 2, 3 and 4 $\sigma$.  The red and cyan contours represent Band\,4 and\,7 continuum data,
          respectively, at 5, 10, and 20 $\sigma$ levels. 
          The beam size for each data set is shown with its corresponding colour (2\farcs{72}\,$\times$\,1\farcs{56}, 0\farcs{85}\,$\times$\,0\farcs{78}, and 0\farcs{30}\,$\times$\,0\farcs{18}, from top to bottom, respectively). 
         }
\label{fig1} 
\end{figure}

\begin{figure*}
	\centering
	\includegraphics[trim= 0 0 0 0, width=180mm]{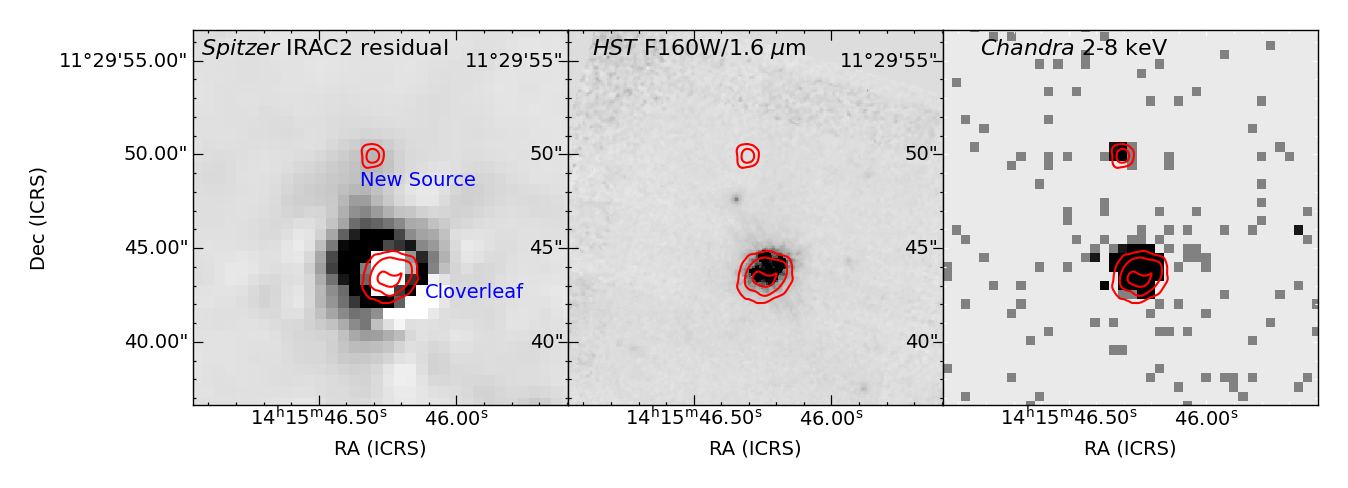}
        \caption{NIR (Left and Middle) and X-ray (Right) snapshots of the new source. 
        The red contours show the ALMA 850\,\mum\ continuum emission, at 5, 10, and 20 $\sigma$. The residual image from {\it Spitzer} 4.5 \mum\ shows marginal detection, while the {\it HST} 1.6 \mum\ band shows nondetection. The new source is detected in the archival {\it Chandra} X-ray data (integrated within in 0.8 -- 10 keV), which originally targeted the Cloverleaf.}
\label{fig2}
\end{figure*} 

\section{ALMA Observations and Data Reduction}
\label{sec:ALMAobs}

\subsection{Band\,3 Observation}
\label{subsec:ALMAband3obs}

ALMA Cycle\,3 band\,3 observations (ID: 2015.1.01309.S; PI: Z.~Zhang) were taken during April 2016, with the pointing centred on the Cloverleaf: right ascension (RA) and declination (Dec.) (J2000) = 14h15m46.25s, $+$11d29m43.4s.  

The observing frequency consists of four spectral windows (SPWs), covering frequency ranges of
91.87--93.74, 93.77--95.75, 103.88--105.86, and 105.77--107.75\,GHz. The synthesized beam is 2\farcs{72}\,$\times$\,1\farcs{56}. 
The exposure time is 29.23\,min, with a precipitable water vacuum (PWV) of 3.12\,mm. 
The details of the observations are described in \citet{zhang2018}. 

\subsection{Follow-up Band\,4 Observation}
\label{subsec:ALMAband4obs}

We performed ALMA band\,4 observations in ALMA Cycle\,7 (ID: 2019.1.00883.S; PI: N.\,H.~Hayatsu). 
The frequency coverage,  136.86--147.81, 148.86--152.60, and 156.06--159.81\,GHz, 
allowed us to look for line emission at possible redshifts: $z = 2.3, 3.4, 4.5$, or $5.6$.  The observation was centred on RA and Dec. (J2000) = 14h15m46.279s, 11d29m49.815s. 

The exposure time is 25.7\,min, with a PWV of $\sim$\,0.75\,mm.  The spectral resolution and the median angular resolution are 1953.09\,kHz and 0\farcs{85}\,$\times$\,0\farcs{78}, respectively. 

\subsection{ALMA Archival Band\,7 Data}
\label{subsec:almaarchive}
We searched the ALMA archive for data targeting the Cloverleaf, finding proposal IDs 2011.0.00747.S, 2012.1.00175.S, 2012.1.01081.S, 2017.1.01232, and 2017.1.00963.S.

In the ALMA Cycle\,1 Band\,7 data (ID: 2012.1.00175.S, PI: P.\,van\,der~Werf), we found the continuum and line emission coinciding with the new source, as described \S\ref{subsec:almaband7}. The exposure time and the angular resolution are 9.4\,min and 0\farcs{30}\,$\times$\,0\farcs{18}, respectively. The frequency resolution is 15.625\,MHz after re-binning, with frequency coverages of 325.84 -- 327.83, 327.28 -- 329.27, 338.41 -- 340.41, and 339.47--341.46\,GHz. 

No significant continuum or line emission was detected in the other archival datasets.

\subsection{ALMA data reduction}
\label{subsec:ALMAreduc}

All ALMA data, including the archival data were reduced and analysed with the Common Astronomy Software Application ({\sc casa}) ver.
5.6.1-8 \citep{mcmullin2007}.  We calibrated the data with the default pipeline scripts. The rms noise 
was estimated using \texttt{imstat} in the region with primary beam response $>0.5$. 
We subtracted the continuum with \texttt{uvcontsub}, then cleaned the data with \texttt{tclean} with briggs weighting and robust $=$ 0.5. 

We fitted a two-dimensional Gaussian profile to the 3 mm continuum data corrected for the primary beam with \texttt{imfit}. Line fluxes are measured with \texttt{specflux}.  The measured values are summarised in Tab.\,\ref{tab1}, combining with the photometry measurements discussed in \S\ref{subsec:conterparts}. There are no significant atmospheric features within the frequency ranges covered by the data used in this paper.

\section{Data Analysis and Source Detection}
\label{sec:analysis}

\subsection{Line Detection in ALMA Band\,3}
\label{subsec:linedetec}

The noise level of the Band\,3 continuum image is $\approx$\,75\,$\mu$Jy\,beam$^{-1}$.  We re-bin the spectral channels from 976.51\,kHz to 15.625\,MHz ($\approx$ 45\kms), which then reaches a noise level of $\approx$\,0.23\,mJy\,beam$^{-1}$ per channel. 

We performed a systematic blind line search in our previous ALMA band\,3 data\,\citep{hayatsu2017}, and discovered a strong emission line at 105\,GHz, coincident spatially with continuum, $\sim$ 6$''$ away from Cloverleaf.
The peak position is RA = 14h15m46.30s, Dec. = 11d29m49.8s. 
The best-fit parameters are: flux density, $218 \pm 75$\,$\mu$Jy, with the major and minor axes of 2\farcs{72} and 1\farcs{72}, respectively, indicating an unresolved source. 

We then extracted a spectrum of the source with an aperture size matched to the beamsize.  As shown in Fig.\,\ref{fig1}, the 105-GHz line profile is doubly or triply peaked, with a full-width zero power (FWZP) $\sim$ 1000 \kms. The two striking peaks lie at 105.00 and 105.22\,GHz, and we take the central frequency as the mean value, 105.11\,GHz. The velocity-integrated flux yields $1.27\pm 0.10$ Jy\,\kms, with the mean of the peaks of $1.73\pm 0.30$ mJy.  The peak SN ratio achieves 6.7\,$\sigma$ at a 45\,\kms\ resolution. The moment-0 image in Fig.\,\ref{fig1} indicates that the line-emitting region is also unresolved.

\subsection{Line Detection in ALMA Band\,4}
\label{subsec:linedetec4}
The sensitivities of the Band\,4 achieves 24\,$\mu$Jy\,beam$^{-1}$ and 0.24\,mJy\,beam$^{-1}$ at a 21.1\,MHz or $\sim$\,40\,\kms resolution, for continuum and line, respectively. 

We measure a continuum flux density of 368\,$\pm$\,70 $\mu$Jy as a point source. 
We extracted a spectrum at the continuum position and the FWZP of the line is $\approx$\,1,200\,\kms, similar to the line found in the band\,3 data. We detected a double or triple peaked emission line at 157.6\,GHz in the band\,4 datacube covering 156.06--159.81\,GHz. The total flux of the line emission is $1.40\pm 0.40$ Jy\,\kms\ with a peak S/N of 8.64\,$\sigma$.

\subsection{Additional Detection in ALMA Band\,7}
\label{subsec:almaband7}
The sensitivity of the Band\,7 continuum achieves 61\,$\mu$Jy\,beam$^{-1}$.  We found a continuum emission at $\sim 340 $GHz, with a flux of $4.0\pm 0.2$\,mJy at the peak position of 14h15m46.31s, 11d29m50.1s with an apparent size of 0\farcs{38}\,$\times$\,0\farcs{36}. 

We obtained the spectrum in the continuum region with the SPWs covering 339.47--341.46 GHz. The median sensitivity of the datacube was $\approx$ 0.6 mJy beam$^{-1}$ at 15.6\,MHz or $\sim$\,13.8\kms resolution.
Weak line emission appears at $\sim  340.8$ GHz. Unfortunately, this SPW covers only half of the necessary velocity range, and only shows a narrow feature.  
The measured total flux of the line is $0.24\pm 0.06$\,Jy\,\kms with a peak S/N of 2.74 $\sigma$.  

\subsection{Photometric Counterparts}
\label{subsec:conterparts}

To check whether the detected lines arise from a known source, we compare the source position of the new source with those of the surrounding galaxies reported by \citet{kneib1998a, kneib1998b}.  We also check the NASA/IPAC Extragalactic Database\footnote{The website of the database:\url{https://ned.ipac.caltech.edu/}}.  There is no known source within a radius of 1$''$ of the position of the new source.  Note that we checked the optically detected absorption line toward the Cloverleaf reported by \citet{monier1998} and there is no corresponding absorption system associated with the newly detected source.

The Cloverleaf field has been observed at multiple wavelengths from radio to X-ray, using, e.g. {\it HST}, the Very Large Telescope (VLT), {\it Spitzer}, {\it WISE}, {\it Herschel}, VLA, ALMA and {\it Chandra}.  We searched for counterparts to the detected source using these archival data, as well as in our newly acquired {\it Spitzer} data (Fig.\,\ref{fig2}). Our measured values and proposal IDs are summarised in Tab.\,\ref{tab1}. In this section, we focus on the nondetections in {\it HST}, VLT, and VLA data and detections in the {\it Spitzer} and {\it Chandra} data.  

\paragraph*{HST and VLT:} 
We perform an aperture photometry for an aperture of $~ 1''$ in diameter for both {\it HST and VLT} data. The {\it HST} multiband optical/NIR data (F814W, F160W, F180M) are obtained from Hubble Legacy Archive\footnote{\url{https://hla.stsci.edu/hlaview.html}}, with project IDs 8268, 7495, and 9439.  The exposure time is 5200\,s, 2560\,s, and 4096\,s for the F814W, F160W, and F180M data, respectively. 
We first masked all strong sources in a 10$''$ radius region, then fit the
background and subtracted it. Then we randomly draw 20,000 times of
$1''$-diameter circles and measure their fluxes. The final noise levels were
obtained by fitting these fluxes with a Gaussian distribution. The nondetections
put upper limits of $< 1.1-2.0$\,$\mu$Jy, corresponding to $\sim$\,23--24 AB
magnitude at $0.8-1.8$\,$\mu$m. 

The VLT/ISAAC data (program ID:67.A-0502; PI: Courbin, F.) in $K$ band shows no detection of the source, to a 3\,$\sigma$ upper limit of $< 2.6$\,$\mu$Jy at 2.3\,\mum. The total exposure time is 1h 50 min. 

\paragraph*{VLA:} 
We obtained the K-a band (33 GHz) JVLA data from the NRAO archival system. The project
ID is 13B-051 (PI: Sharon, Chelsea), and was taken from 27 October 2013 to 5
January 2014, with the B-array configuration. The total on-source time is
$\sim$ 6.8 hours on-source. 
The data were calibrated and imaged with CASA 5.6.2-2 using the standard pipeline. 
The noise level is $\sigma \sim 4.5\times 10^{-6}\  \rm{Jy\ beam^{-1}}$, with a beam size of $\sim 0.''25$.  

\paragraph*{Spitzer/IRAC:} We obtained {\it Spitzer} observations with 500\,s of on-source time in IRAC band\,1 (3.6 \mum), and 1080\,s in the IRAC band\,2 (4.5 \mum), with the proposal ID:14252 (PI: N.\,H.\,Hayatsu).  This source was observed on 28 October, 2019.

The observation yielded an rms noise level of 0.39 $\mu$Jy at 3.6\,\mum\ and 0.42 $\mu$Jy at 4.5 $\mu$m. 
Our target is covered and overpowered by the halo of the Cloverleaf because of the large PSF of IRAC images.  The residual IRAC image was generated after the subtraction of the contribution of the Cloverleaf.  The IRAC subtraction was performed using the {\it HST}/F180M image as a model for Cloverleaf. We convolve the point sources in the F180M image with the IRAC PSFs rotated to the observing position angle to make the Cloverleaf model in both IRAC images. The aperture corrections were applied. 

After removing the Cloverleaf using the models, we performed aperture photometry on the location of the new source. We adopted the sky subtraction value from the further out region (with a $\sim 20''$ annulus radius) and the mean statistics clipped to avoid contamination from the residual of the Cloverleaf. The error was estimated by combining the sky variation within the sky annulus and the residual variation near the new source. We also propagate the calibration errors into the error estimate. 

The source was detected in the residual map with flux densities of $31\pm 7$\,$\mu$Jy and $38\pm 6$\,$\mu$Jy for the IRAC band\,1 and band\,2, respectively.

\paragraph*{Chandra:} We have identified the X-ray counterpart from the archival {\it Chandra} data targeting at Cloverleaf with a total exposure time of 147\,ks after removing 38.2\,ks of exposure affected by a solar flare in April 2000. We extracted Chandra X-ray spectra from a 1$''$ radius circular region centered at the new source, and selected the background from an annular region surrounding the new source. By merging all the observations, Our analysis yields 45 counts of X-ray photons in 0.5--10\,keV, which were grouped to have a minimum of 5 counts per bin.  There was no positional offset between the X-ray and submm dust continuum emission from ALMA (within $0.''8$). 

\begin{table}
\caption{Position, line total flux, and multi-band photometry of the new source.}
\begin{threeparttable}
\begin{tabular}{rcl}
\hline 
Parameter & Value & Comment or unit\\
\hline \hline
R.A.& 14:15:46.279 & ALMA, J2000\tnote{(1)}\\ 
Dec & 11:29:49.815 & ALMA, J2000\tnote{(1)}\\
redshift & $3.386 \pm 0.005$ & from CO $J$\,=\,4--3\tnote{(1)} \\\hline
CO $J=4$--3 & $1.27 \pm 0.10$ & Jy \kms\tnote{(1)}\\
CO $J=5$--6 & $1.40 \pm 0.40$ & Jy \kms\tnote{(2)}\\
CO $J=12$--11 & $0.24 \pm 0.06$ & Jy \kms\tnote{(3)}\\ \hline
0.5--10 keV & 45 & counts, {\it Chandra} \tnote{(4)}\\
0.8 \mum\ & $< 2.0$ & $\mu$Jy,{\it HST}/F814W\tnote{(5)}\\
1.6 \mum\ & $< 1.2$ & $\mu$Jy,{\it HST}/F160W\tnote{(6)}\\
1.8 \mum\ & $< 1.1$ & $\mu$Jy,{\it HST}/F180M\tnote{(7)}\\
2.3 \mum\ & $< 2.6$ & $\mu$Jy, VLT/ISAAC $K_{\rm s}$\tnote{(8)}\\ 
3.6 \mum\ & $31 \pm 7$ & $\mu$Jy, {\it Spitzer}/IRAC1\tnote{(9)}\\
4.5 \mum\ & $38 \pm 6$ &  $\mu$Jy, {\it Spitzer}/IRAC2\tnote{(9)}\\
340 \,GHz & $4.0 \pm 0.2$ & mJy, ALMA/band\,7\\
140 \,GHz & $368 \pm 70$ & $\mu$Jy, ALMA/band\,4\\
100 \,GHz & $218 \pm 75$ & $\mu$Jy, ALMA/band\,3\\
33 \,GHz & $< 15$ & $\mu$Jy, VLA \tnote{(10)}\\ \hline
\end{tabular}
\begin{tablenotes}
\item[] Note: All upper limits are 3\,$\sigma$.
\item[(1)] ALMA band\,3 data (2015.1.01309.S).
\item[(2)] ALMA band\,4 data (2019.1.00883.S).
\item[(3)] ALMA band\,7 data (2012.1.00175.S). 
\item[(4)] IDs: 930, 5645, and 14960.
\item[(5)] The ID is 8268 and the instrument is WFPC2/PC.
\item[(6)] The ID is 7495 and the instrument is NICMOS.
\item[(7)] The ID is 9439 and the instrument is NICMOS.
\item[(8)] Co-add data.
\item[(9)] ID:14252.
\item[(10)] ID:13B-051.
\end{tablenotes}
\end{threeparttable}
\label{tab1}
\end{table}

\section{Result and Discussion}
\label{sec:result}

\subsection{Redshift Identification}
From the frequency ratios of the detected line emission, we identify the line species are CO $J$ = 4--3, 6--5, and 13--12 for the 105.1, 157.6, and 340.8\,GHz lines, respectively. Therefore, we obtain the redshift of this system, from the first detection at $105.11\pm 0.11$,GHz, corresponding to $z=3.386 \pm 0.005$.  Note that other redshift possibilities from $z = 0$--10 are excluded by the nondetection of line emission in our other observed frequency range. 

\subsection{CO Emissions and Molecular Mass} 
The line width, full width at zero power (FWZP) $\sim$ 1000 \kms, and the double-peaked CO $J$ = 4--3 and 6--5 lines, are suggestive of an interacting system \citep[e.g.,][]{salome2012, treister2012, diaz2018}. The frequency coverage of CO $J$=13--12 line covers only part of the low-$J$ velocity range (Fig.~\ref{fig1}).  Note that alternative explanations for the double-peaked emission profile, such as a disk-like galaxy, should be investigated in future high-angular resolution observations.

We estimate its molecular gas mass $M_{\rm mol}$ from the CO $J$= 4--3 line using conversion factors for mass-to-light ratio and CO line ratio.  We use a formula from \citet{bolatto2013}: 
\[
M_{\rm mol} = 1.05 \times 10^4 \left( \frac{X_{\rm CO (1-0)}}{2 \times 10^{20} \frac{\rm cm^{-2}}{\rm K~km~s^{-1}}} \right)\frac{S_{\rm CO(1-0)} \Delta v D_{\rm L}^2}{(1+z)}~~{\rm M_\odot}, 
\]
where $X_{\rm CO}$, $S_{\rm CO}\Delta v$ $D_{\rm L}^2$, and $z$ are CO-to-H$_2$ conversion factor in cm$^{-2}$ K$^{-1}$ km$^{-1}$ s, integrated flux density for CO $J$=1--0 in Jy km s$^{-1}$, luminosity distance in Mpc, and redshift. 

We assume a $X_{\rm CO(1-0)} = 0.5 \times 10^{20}$cm$^{-2}$ K$^{-1}$ km$^{-1}$ s, a typical value for high-redshift starburst galaxies \citep{bolatto2013}. 
We refer to the high-$z$ CO $J$ = 4--3/CO $J$ = 1--0 line brightness temperature ratio $r_{41} = 0.63 \pm 0.44$ from \citet{frias2023}, assuming the same line profile between the two CO line emissions.  

Although the value differs by the selection of the conversion factor and the assumption of the line profile, the estimate indicates that the source has a massive molecular gas budget $M_{\rm mol}$ $\sim$ 4--23 $\times 10^{10}$ M$_\odot$, which is consistent with the typical value of high-$z$ ULIRGs. 

\subsection{X-ray Emission and BH mass}
To fit the X-ray spectrum, we first applied an absorbed power-law model \texttt{zpowerlaw}, \texttt{tbabs}, and \texttt{ztbabs} in Xspec \citep{xspec1996} and employed a C-statistic \citep{Cstatics1979} for fitting statistics. 
The redshift is fixed to 3.39 and the photon index is fixed $\Gamma = 1.7$, a typical value for AGN. 

The X-ray spectrum suggests an intrinsic absorption column density $N_{\rm H}$ of $(5\pm 2) \times 10^{23}$~cm$^{-2}$, which consistent with the estimate from the ALMA dust measurement discussed in Appendix \ref{sec:SED}.  X-ray flux are estimated $4.6\times 10^{-15}$ erg\,cm$^{-2}$\,s$^{-1}$ and the total X-ray luminosity is $\sim 4\times 10^{44}$ erg\,s$^{-1}$ in 0.5–10 keV.  From this value, we can obtain an estimate of black hole mass $M_{\rm BH} \sim 10^8 {\rm M}_\odot$ from the empirical correlation reported in \citet{mayers2018}. 
The hard X-ray data show a point 
source, suggesting a dust-enshrouded AGN (Fig.~\ref{fig2}).  

The nondetection at the optical wavelength indicates that this system, no matter whether 
it is powered by starburst or AGN, is highly obscured.
Although the nondetections at the optical wavelength suggest that the AGN is obscured \citep[e.g.,][]{wang2019}, the X-ray detection indicates that the foreground material is not Compton thick for X-ray. It is suspected that the AGN has not yet reached its peak accretion capacity and is not fully cocooned by dust. The molecular gas has not yet been impacted by radiative feedback. This scenario is similar to that of extremely luminous Hot DOGs in a similar redshift range \citep{tsai2015}, although these Hot DOGs usually reach the Eddington limits \citep{wu2018, tsai2018} and lack AGN features in their optical wavebands and X-ray emissions \citep{vito2018}. Therefore, it is possible that the new source is in the progenitor stage of the hot DOG phase, perhaps evolving towards the naked AGN phase and then the elliptical galaxy in the local universe \citep{thomas2010, toft2014}.  

We present SED analysis and stellar mass estimation in Appendix \ref{sec:SED}.

\section{Summary}
\label{sec:summary}
We discovered a new galaxy located $6''$ away from the well-known Cloverleaf quasar (H1413+117) at $z=2.56$. The spectroscopic redshift of this system was determined to be $z=3.386 \pm 0.005$ based on detections of the CO $J$ = 4--3, 6--5, and 13--12 lines from ALMA band 3, 4, and 7 data, respectively.

The double-peaked CO profiles suggest that the molecular gas in this system has not yet settled dynamically, likely due to a gas-rich major galaxy merger in its early stages. Therefore, the source is likely destined to continue as a starburst, with its rich molecular budget, triggered by the merging process \citep[e.g.,][]{glikman2015}. In addition, the strong X-ray emission provides evidence of an accreting AGN in this system.  

The central mystery of this source compared to typical SMGs is how it can display bright dust continuum and X-rays alongside extremely weak NIR/optical. The newly confirmed system at $z = 3.39$ is thus an interesting target to study galaxy--BH co-evolution in extremely dusty conditions. To infer the evolutionary stage of such an optically dark galaxy, be it massive starburst, QSO, or HotDOG, it is essential to clarify the physical state of the gas, especially the dense molecular gas that provides the fuel for star formation, and this will be the subject of future follow-up studies.  The uncertainty of the mass estimates should be discussed and improved after obtaining new constraints for NIR and FIR wavelengths using e.g., JWST and ALMA.

\section{acknowledgements}
  We would like to express our sincere gratitude to the anonymous reviewers for their valuable comments and suggestions, which significantly improved the manuscript.
  
NHH acknowledges the support of European Southern Observatory (ESO) through its PhD Studentship Program. NHH is grateful to the NAOJ ALMA project for financial support for the ALMA workshops (2020b, 2023a). This work was initiated and conducted during NHH's time at ESO, the University of Tokyo, Bunkyo University, and Ecole Normale Sup\'erieure, and was submitted while she was affiliated with the Ishigakijima Astronomical Observatory, NAOJ.

  Z.-Y. Zhang acknowledges the support of NSFC (grants 12041305, 12173016, 1257030642, 12533003).
  Z.-Y. Z. acknowledges the Program for Innovative Talents, Entrepreneur in Jiangsu. 
  Z.-Y. Z.  acknowledges the science research grants from the China Manned Space Projects with NO. CMS-CSST-2021-A08, and CMS-CSST-2021-A07. 
  Funded by the Deutsche Forschungsgemeinschaft (DFG, German Research
  Foundation) under Germany's Excellence Strategy -- EXC-2094 --
  390783311.
  
  C.-W. Tsai acknowledges support from the NSFC grant 11973051.
  
  C. Yang acknowledges support from an ERC Advanced Grant 789410.
  
  D. Burgarella acknowledges the CNRS International Research Network France-Japan NECO".
  
  This work was supported by JSPS KAKENHI Grant Numbers JP18H05437, JP20H01950 (MN), JP17H06130, JP23K20035 and JP24H00004 (KK), 
  and the NAOJ ALMA Scientific Research Grant Number 2017-06B (YN, KK).  
  
  J. Wang acknowledges support from the NSFC grant 12033004. 
  
  This paper makes use of the following ALMA data:
  ADS/JAO.ALMA\#2011.0.00747.S, 2012.1.00175.S, 2015.1.01309.S,
  2017.1.00963, 2017.1.01081.S, 2017.1.01232.S and
  2019.1.00883.S. ALMA is a partnership of ESO (representing its
  member states), NSF (USA) and NINS (Japan), together with NRC
  (Canada), MOST and ASIAA (Taiwan), and KASI (Republic of Korea), in
  cooperation with the Republic of Chile. The Joint ALMA Observatory is operated by ESO, AUI/NRAO, and NAOJ.
  This research has made use of the NASA/IPAC Extragalactic Database
  (NED), which is funded by the National Aeronautics and Space
  Administration and operated by the California Institute of
  Technology.
  
\section{Data Availability}
  The data underlying this article were accessed from
  \begin{itemize}
    \item[] {\it HST}~: \url{https://hla.stsci.edu/hlaview.html}, 
    \item[] VLT~: \url{https://archive.eso.org/scienceportal/home}
    \item[] VLA~: \url{https://data.nrao.edu/portal/#/}
    \item[] {\it Spitzer}~:\url{https://sha.ipac.caltech.edu/applications/Spitzer/SHA/}
    \item[] {\it Chandra}~: \url{https://cda.harvard.edu/chaser/}
    \item[] ALMA~: \url{https://almascience.nrao.edu/aq/}. 
  \end{itemize}
   The derived data generated in this research will be shared on reasonable request to the corresponding author. 


\bibliographystyle{mnras}
\bibliography{DV_refs} 

\appendix
\section{SED properties}
\label{sec:SED}

\begin{figure*}
\centering
\includegraphics[trim= 0 10 10 0, width=170mm]{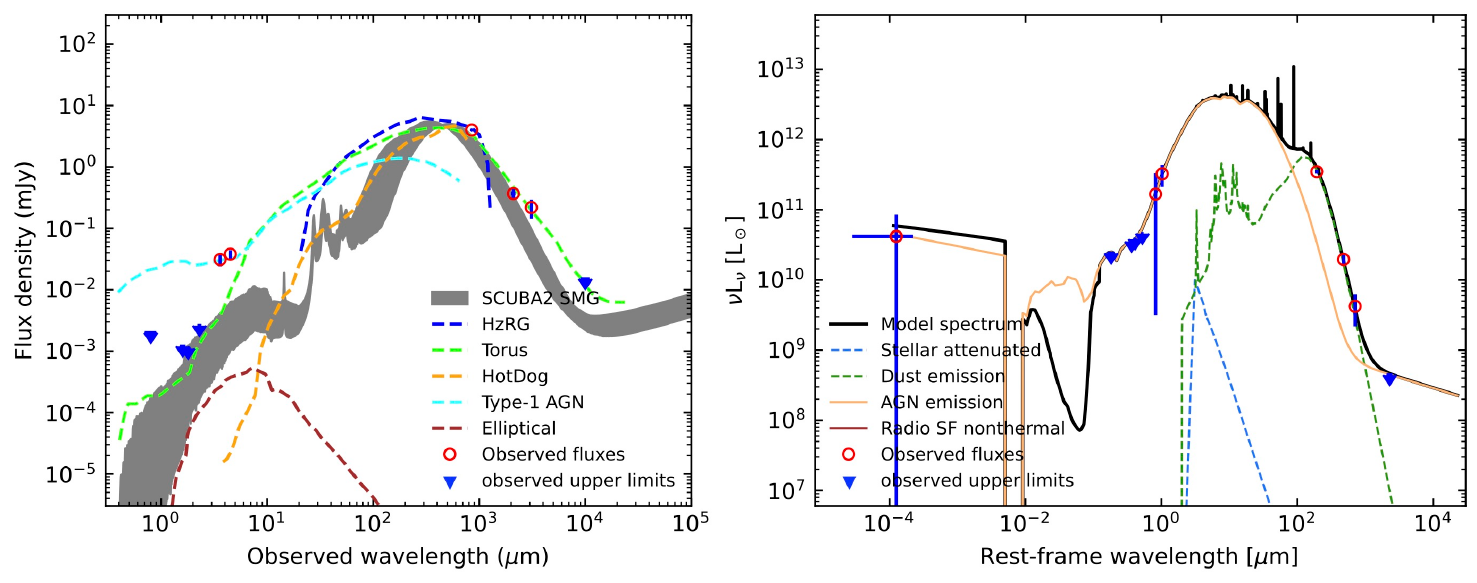}
\caption{Photometry of the new source. The measurements are shown in red circles and blue triangles. {\it (Left:)} Blue dashed, light green dotted, orange, cyan, and brown curves represent the SED templates of high-redshift radio galaxy, DKB07 \citep[HzRG,][]{Dannerbauer2014},
AGN torus model constructed from Compton-thick AGNs \citep{polletta2006}, hot dust-obscured galaxies \citep{tsai2015}, a Type-1 AGN \citep{richards2006}, and an elliptical galaxy \citep{assef2010}, respectively.  The grey shaded region indicates the 1-$\sigma$ region for a large sample of SMGs from SCUBA-2 \citep[following][]{ivison2019, Duzeviciute2020}. None of the templates are compatible with the SED of the new source. {\it (Right:)} The best fit model from the SED fitting code X-CIGALE \citep{Cigale, Cigale2, Xcigale}.  The best-fitting model well explains the upper limits in the optical wavebands (black solid curve). The dashed lines show the corresponding SED components.}\label{fig:sed}
\end{figure*}

We compile the SED of the new source with all data collected from the literature (see Fig.~\ref{fig:sed} left). 
These data cover a wide range of wavelengths, from radio to optical. We then attempt to infer the global
physical properties of the new source by comparing with a series of SED templates of a high redshift radio galaxy \citep[HzRG][]{Dannerbauer2014}, an AGN torus model constructed
from Compton-thick AGNs \citep{polletta2006}, hot dust-obscured galaxies \citep{tsai2015}, a Type-1 AGN \citep{richards2006}, an elliptical galaxy \citep{assef2010},  and SCUBA-2 SMGs \citep{Duzeviciute2020, ivison2019}, respectively. We manually adjusted the amplitude of the template to best match the observed data.

Overall, none of the SEDs is compatible with the optical darkness of the new source. This suggests that 
the emission in our deep {\it Spitzer} IRAC measurements is dominated by the torus emission constructed 
from a Compton-thick AGN \citep[light green dotted curve in Fig.~\ref{fig:sed} left][]{polletta2006}, but yet still 
suffers extinction which removes most of its rest-frame optical emission. The nondetection by {\it HST} 
naively suggests a host galaxy with $M_*$ smaller than that of elliptical galaxy (10$^{10}$ M$_\odot$; the brown curves shown in Fig.~\ref{fig:sed} left), but of course this ignores the effects of dust obscuration.  

For more discussion, we use SED fitting code CIGALE (Code Investigating GALaxy Emission\footnote{\url{https://cigale.lam.fr}}), which is capable of modelling the observed SED from X-ray to radio \citep{Cigale, Cigale2, Xcigale} and returns the best-fit model and the Bayesian-like analysis.  We aim to obtain far-infrared luminosity ($L_{\rm FIR}$), stellar mass, and AGN fraction from the result.  We input fixed values for, e.g., redshift, X-ray power-law index so that the values are consistent with the observational result and the assumption. For some of the other parameters, such as visual extinction $E(B-V)$, or a fraction of the AGN luminosity to the total dust bolometric luminosity $f_{\rm AGN}$, we input as a range of values. For example, we input $E(B-V)$ = 30--50 assuming that the column density of HI is the same as the column density from X-ray observation. 

We run $\sim 3 \times 10^7$ models and the best model which contains six components; an attenuated stellar, a nebular, a dust, an AGN, and a radio component. For detailed information on the parameter sets, see Tab.\,\ref{tab2}. Note that we ignore the unreal cuts at the edge of the components. The best-fitting model with reduced $\chi^2$ = 3.55 explains the upper limits in the optical wavebands  (black solid curve in Fig.\,\ref{fig:sed} right). 
As the results, the Bayesian analysis yields large uncertainties in model parameters. However, we still consider the results for the rough estimates on the overall characteristics of the SED such as the luminosity or mass.

From the best fitting model, we obtain $M_* = 5.4\pm 0.8 \times 10^{10}$\,M$_\odot$. 
This large uncertainty indicates the model cannot constrain stellar mass. Note that this cannot be improved with the higher extinction ($E(B-V) > 100$). Thus we regard the value $M_* \lesssim 10^{11}$\,M$_\odot$ as an upper limit. 

The dust bolometric luminosity is $L_{\rm FIR}$ = $2.8\pm2.3$ $\times 10^{12}$\,L$_\odot$. The value from the best-fit model is lower than a conservative estimate from the dust continuum flux, $\sim$ 10$^{13}$ L$_\odot$ \citep[e.g.,][]{blain2002}. 

The extinction and the AGN fraction $f_{\rm AGN}$ shows $40\pm8$ and $0.5 \pm 0.3$ for the Bayesian-like analysis, respectively. 

To compare the star-formation rate (SFR) with the known galaxies, we obtain SFR by considering the contribution from dust bolometric luminosity using the empirical relation from \citet{kennicutt1998} assuming \citet{chabrier2003} IMF, which is the same assumption to the SED analysis:
\[
\frac{\rm SFR}{\rm M_\odot yr^{-1}}  \approx 1.52 \times 10^{-10} x_{\rm StoC} (1-f_{\rm AGN}) ~\frac{L_{\rm FIR}}{\rm L_\odot},
\]
where the conversion factor from Salpeter to Chabrier IMF $x_{\rm StoC} = 1.7$.  The discovered source has an SFR $\sim 362^{+693}_{-336}~{\rm M}_{\odot}{\rm yr}^{-1}$. The SFR from Cigale returns $\sim$ 83 $~{\rm M}_{\odot}{\rm yr}^{-1}$, which is consistent with the calculation.
The dust bolometric luminosity requires the contribution of an SMBH with $M_{\rm BH}$ = 1.4$^{+2.6}_{-1.3}\times 10^8$ M$_\odot$, assuming an typical Eddington ratio $\sim$ 0.3 for $z \sim$ 2 QSOs \citep{assef2015,tsai2015}.  
The estimate has an agreement with $M_{\rm BH}$ value from X-ray. 

\begin{table*}
\caption{A range of models, free parameters and best-fitting model for the new source's SED, calculated by X-CIGALE.}
\begin{threeparttable}
\begin{tabular}{rcc}
\hline 
Model/Parameter & Values & Best-fitting model\\
\hline \hline \\
\multicolumn{3}{c}{\bf Star-formation history: double-exponentially decreasing ($\tau$-decay) model}\\ \hline
e-folding time of the main stellar population (Myr) & 500, 1000, 2000, 3000, 6000 & 6000 \\
Age of the main stellar population (Myr) & 100, 250, 500, 1000, 2500, 5000 & 1000 \\
e-folding time of the late starburst population (Myr) & 25, 50, 75, 100 & 50 \\

Age of the late burst (Myr) & 5, 10, 20, 50, 100 & 20\\ 
Mass fraction of the late burst population & 0.01, 0.03, 0.1 & 0.01 \\ \hline
\\
\multicolumn{3}{c}{{\bf Stellar population synthesis model}}\\ \hline
Simple stellar population & Bruzual \& Charlot (2003) & \\ 
Initial mass function & Salpeter (1955) or Chabrier et al. (2003) & \citet{chabrier2003}\\
Metallicity & 0.02 &  0.02 \\ \hline 
\\
\multicolumn{3}{c}{{\bf Dust attenuation: Calzetti et al. (2000)}}\\ \hline
Colour excess of stellar continuum light for young stars E(B-V) & 30 -- 60 (step 5) & 50 \\ \hline
\\
\multicolumn{3}{c}{\bf Dust emission: Dale et al.(2014)}\\ \hline
$\alpha$ slope in d$M_{\rm dust} \propto U^{-\alpha}$d$U$ & 1.0, 2.0, 3.0 & 2.0 \\
(IR power-loaw slope) & & \\ \hline
\\
\multicolumn{3}{c}{{\bf AGN (UV-to-IR): SKIRTOR (Stalevski et al. 2012, 2016)}}\\ \hline
Viewing angle ($\theta$) & 30\textdegree, 30\textdegree & 30\textdegree\\ 
(face on: $\theta$ = 0\textdegree, edge on: $\theta$ = 90\textdegree) &  & \\
Opening angle & 20\textdegree, 50\textdegree, 70\textdegree & 50\textdegree\\
a AGN fraction in total IR luminosity & 0.1, 0.3, 0.5, 0.7, 0.9 & 0.9\\
Mass fraction of the late burst population & 0.4, 0.5, 0.6 & 0.6 \\
Extinction law of polar dust & SMC, Calzetti, Gaskell2004 & SMC\\
E(B - V) of polar dust & 30, 40, 50 & 50\\
Temperature of polar dust (K) & 100, 150, 200 & 100\\
Emissivity of polar dust & 1.6 & \\ \hline
\\ 
\multicolumn{3}{c}{{\bf AGN X-ray}}\\ \hline
AGN photon index ($\Gamma$) & 1.7 &\\ 
X-ray energy cut-off ($E_{\rm cut}$) & 330 KeV &\\
Maximum deviation from the $\alpha_{\rm ox}$-$L_{\nu}$(2500 $\AA$) relation& 0.2 & \\ \hline
\\
\multicolumn{3}{c}{{\bf Radio}}\\ \hline
The slope of the power-law synchrotron emission & 0.7, 0.8, 0.9 & 0.9  \\  \hline
\end{tabular}
\end{threeparttable}
\label{tab2}
\end{table*}


\bsp	
\label{lastpage}
\end{document}